\documentclass[sigconf]{acmart}
\usepackage{multicol}
\usepackage{tikz}
\usepackage{algorithm}
\usepackage{float}
\usepackage{subfig}
\usepackage[most]{tcolorbox}
 \usepackage{algorithm}
 \usepackage{algpseudocode}

\begin{document}

\title{A Geometric Approach for Real-time Monitoring of Dynamic Large Scale Graphs}
\subtitle{AS-level graphs illustrated.}

%
%
%
\author{Loqman Salamatian}
\affiliation{%
\institution{CSIRO Data61}
\city{Sydney}
\state{Australia}
}
\email{salamatianloqman@gmail.com}
\author{Dali Kaafar}
\affiliation{%
\institution{Macquarie University/CSIRO Data61}
\city{Macquarie}
\state{Australia}
}
\email{Dali.Kaafar@data61.csiro.au}
\author{Kav\'e Salamatian}
\affiliation{%
\institution{Universit\'e de Savoie}
\city{Annecy}
\state{France}
}
\email{kave.salamatian@univ-smb.fr}



\begin{abstract}
The monitoring of large dynamic networks is a major challenge for a wide range of application. The complexity stems from properties of the underlying graphs, in which slight local changes can lead to sizable variations of global properties, {\em e.g.}, under certain conditions, a single link cut that may be overlooked during monitoring can result in splitting the graph into two disconnected components. Moreover, it is often difficult to determine whether a change will propagate globally or remain local. Traditional graph theory measure such as the centrality or the assortativity of the graph are not satisfying to characterize global properties of the graph.
In this paper, we tackle the problem of real-time monitoring of dynamic large scale graphs by developing a geometric approach that leverages notions of geometric curvature and recent development in graph embeddings using Ollivier-Ricci curvature \cite{Ollivier:2009}. We illustrate the use of our method by considering the practical case of monitoring dynamic variations of global Internet using topology changes information provided by combining several BGP feeds. In particular, we use our method to detect major events and changes via the geometry of the embedding of the graph. 

We first adapt the Ricci curvature to characterize the AS level graph. The key idea is to detect changes in between consecutive snapshots and to separate events that result in considerable geometric alterations, from those that remain local. The variations of curvature are evaluated through a set of landmarks as reference points. We develop an anomaly tracking mechanism to detect large variations of curvature that translate into major variations in the geometry of the graph.
These changes are then considered as major BGP level events. In a second stage, we describe a mechanism for identifying the network elements responsible for the set of coordinated changes and isolate their geometric implications. We evaluate this system in operational settings and show its performance in real-life scenarios.
\end{abstract}
\acmConference[IMC 2018]{}{Nov. 2018}{}
\setcopyright{acmcopyright}
\acmISBN{ISBN 978-x-xxxx-xxxx-x/YY/MM}
\acmDOI{nnnnnnn.nnnnnnn}

\acmConference[IMC 2018]{}{Nov. 2018}{}
\setcopyright{acmcopyright}
\acmISBN{}
\acmDOI{}


\maketitle

\section{Introduction}
Monitoring large scale dynamic graphs such as the Internet, neuronal connections or social networks is a challenging task, mainly due to their rapidly changing properties. In particular, slight local changes in a graph, such as a link removal, can result in major variations in global properties. For instance, a few disconnected components may yield great changes in the average lengths between vertices. It is often challenging to distinguish changes in a graph that have only a local impact, from those that propagate globally.



The relevance and difficulty of the efficient monitoring of large scale networks is illustrated in the monitoring of global Internet through information provided by the Border Gateway Protocol (BGP). BGP is the standard inter-domain routing protocol used over Internet to primarily  exchange reachability information among Border routers in Autonomous Systems (AS). 
Dynamic variations of the network resulting from local events, {\em e.g.}, link failures, router reboot, network operator policies changes or  malicious events may lead to a continuous stream of prefix updates or removal. 
The stream that reflects localized changes in the topology of Internet can be observed in the routing tables updates. While the impact of these events  remains generally localized on the affected ASes ({\em e.g.} an outage in a stub AS might disconnect it from Internet and perturb only direct communications) sometimes the reach is larger and the impact deeper. In August 2017, a BGP leak \footnote{illegitimate advertisement for prefixes generally due to misconfiguations} from a single stub AS, (one of Google's in this case), led to a partial black-out in Japan, resulting into a major and large scope disruption \cite{japan-event}. These larger scope events might affect multiple ASes, largely beyond the ones directly involved in the source BGP event, in particular with how widespread BGP security risks are \cite{Survey-BGP}.

Several methods for detecting and mitigating the impact of BGP-level network events such as \emph{BGP hijacks} have been introduced. Proactive mechanisms, like RPKI \cite{wahlisch}, rely on cryptography mechanisms to authenticate the source of BGP announcements and the sequence of updates \cite{Butler}.
Other approaches are reactive and involve first detecting that a suspicious (or abnormal) event has happened and then taking circumvention actions \cite{Kruegel}\cite{Karlin}. The detection of BGP related events is often implemented by third-parties that monitor several BGP vantage points and have therefore a broader view of the BGP activity \cite{caida}\cite{Orsini}. These approaches leverage the characteristics of BGP hijacks as seen in practice, and combine the views from several vantage points (up to 100 collection points for some of them) to implement heuristics to detect BGP events that are then made available to AS owners \cite{dyn}.

While these approaches might be successful identifying the source of the abnormal event, they are unfortunately incapable of tracking the magnitude of an event within the BGP-derived AS graph and of characterizing its importance and global impact. In this paper, our aim is to develop a monitoring system that detects quasi-instantaneously whether important changes have happened in the network and determine how they may have affected the network. 

Understanding and characterizing the scope of BGP events is important both for security purposes and for troubleshooting \cite{Li}\cite{Labovitz}. We consider two application settings to motivate our method. In the first setting we consider an AS, possibly multi-homed, that would like to monitor large scale changes in the Internet connectivity. In the second setting, we consider a third-party actor, possibly a governmental authority, that is interested in monitoring the global stability of Internet and in this capacity would like to detect major changes on Internet connectivity. We will show later that the approach developed in the paper is relevant to both scenarios.

\noindent \textbf{Overview of the technique:} At each snapshot, the network can be defined as a graph with a set of vertices linked through edges. Graph embedding techniques \cite{Cai2018} can be applied to embed this graph into a space endowed with a metric capturing the interrelation between local and global connectivity.  The motivation for embedding graphs in manifolds comes from results in continuous manifold theory, namely the Gauss-Bonnet theorem \cite{Lafontaine2015} which relates a local property, the curvature, to global topological properties of the continuous manifold, {\em e.g.} number of holes. The Gauss-Bonnet theorem ensures that an increase in curvature in one part of the manifold, has to be compensated by a decrease in another part, unless an important change in the topology has happened, {\em i.e.} the manifold acts as a play-dough that when squeezed somewhere will inflate somewhere else. 

In this context we use a discretization of the Ricci curvature \cite{Ollivier}. The Ricci curvature has been widely used in continuous manifold theory to obtain volume and propagation inequalities such as the Brunn-Minkowski theorem \cite{brunn}, and relates the local properties of a manifold with its overall shape. The Ollivier-Ricci curvature defines an extension of the Ricci curvature in discrete settings and is defined through the notions of {\em optimal transport} and {\em transport plan}.  The optimal transport formalizes the problem of finding (if it exists) a function that transforms a distribution of weight over a set of vertices into another desired distribution, while minimizing the product of the amount of transported weight with the distance to which it was transported \cite{levy}. We provide further details in Sections \ref{sec:RiemanianManifolds} and \ref{sec:ricci}. More in depth descriptions can also be found in \cite{Lafontaine2015, Ollivier}.

By projecting every element of a graph $G = (V,E)$ with $|V| = N$ vertices with respect to the Ollivier-Ricci curvature, we obtain a manifold located in $\mathbb{R}^N$ where the coordinates of vertex $x$ are the Ollivier-Ricci curvatures $\kappa(x,y)$ to all other vertices $y \in V$. A dynamic network can, therefore, be seen as a drifting process where vertex positions are fluctuating over time. This embedding of the graph in $\mathbb{R}^N$ is called the Ollivier-Ricci graph Embedding (O.R.E.). In practice, however, embedding the graph in a $N$  dimensional space is a daunting task. To reduce the dimensionality, we define $K$ landmarks that act as anchors in the space, and we project the graph into a $K<<N$-dimension space. We will present a way of selecting the landmark in a way that the loss of information is minimal. 




Assuming we know this embedding, our goal is to detect major changes in the geometry and topology of the BGP graph. When a local change propagates globally within a manifold, it impacts the overall curvature, {\em i.e.}, it changes a large portion of the shortest paths on the manifold (named geodesics).


\noindent \textbf{Main contributions : } 
The contributions of this paper are as follows:
\begin{enumerate}
    \item We propose in section~\ref{sec:ricci}, a geometric view-point to the analysis of AS-level graphs as measured through multiple BGP collectors. We introduce the background material briefly describing the Riemannian geometry and present metrics applicable to AS-level graphs. We use the Ricci curvature and its extension, the Ollivier-Ricci Curvature to characterize the geometry of discrete graphs, in particular the AS-level graphs. We show that the Ollivier-Ricci \cite{Ollivier:2009} curvature can be used as a metric to quantify the impact of a local change on the geometry of the network. These changes can be interpreted via the so-called optimal transport, thus giving a detailed picture of how a change will affect the global network traffic. This provides a new way of quantifying the severity of BGP events. 
     \item We tackle, in section ~\ref{sec:landmarks}, the problem of finding landmarks in highly dynamic settings. We provide and validate a statistical approach to discover optimal landmark points.
    \item We develop, in section \ref{sec:Ollivier}, a monitoring system that uses the Ollivier-Ricci curvature to evaluate if incoming BGP updates have resulted in geometric changes over the global Internet. 
    \item We evaluate, in section ~\ref{sec:evaluation}, this geometry change detection framework over real BGP flows and show that the monitoring system detects all known major BGP-related events of the past years. We also validate that local events can be distinguished from global events. During the evaluation, we compare the Ollivier-Ricci embedding metric with the traditional metrics (the shortest path distance or the spectral distance \cite{Luxburg}). We show that the Ollivier-Ricci distance exhibits better stability in events associated to local changes while showing great sensitivity for events of global changes. We also show that the Ollivier-Ricci metric can uncover relationships between ASes that are invisible to other metrics.
    
\end{enumerate}

\section{Background}

\subsection{BGP collection and AS level graphs}
\label{sec:BGP}

Interdomain routing is a collaborative effort among ASes, which interconnects and exchanges routing information using the BGP protocol. Two main types of announcements messages are exchanged between routers running the BGP protocol: updates that contain each an AS path composed of the list of ASes to cross in order to reach a given prefix through the announcing neighboring AS, and withdrawals announcing that a IP address prefix is not anymore reachable through the announcing AS \cite{Caesar}. 

The decision to announce a path depends on route export policies that derive from contractual relationship between ASes as well as traffic engineering implemented by ASes to manage the traffic exchanges.  Generally, both ASes relationships and traffic engineering policies are maintained confidential. 

Any BGP router will have to deal with a continuous flow of update or withdrawal messages that reflect every slight change in the network topology. When receiving an update for a prefix, a router check in his routing table if it already knows an AS path to it. If this is the case, it will choose between the known active AS path and the new incoming one \cite{Bonaventure}. This choice can be done using the AS path length, {\em i.e.}, a shorter AS path length being better, or through more complex mechanism involving {\em local preferences}, {\em multi-exit-discriminator} path attributes or AS policies. An updated path might be re-announced to neighbors or rather filtered depending on the AS policies. Sometimes AS path are inflated by repeating one of several ASes to make them less likely to be chosen as forwarding path as they become long \cite{Zhang}. 

Another characteristic of BGP is its lack of authentication for routes announcement. It is therefore possible for an AS able to advertise illegitimately paths for prefixes it does not own. These illegitimate advertisements, called BGP Prefix hijacks, can impact many ASes, or even the whole Internet, affecting service availability and security of communications as packets might follow paths they are not supposed to take, {\em e.g.}, because surveillance is implemented along this path. BGP hijack might also be caused by router misconfiguration or malicious attacks \cite{Ballani}.

When an AS path is announced, with two consecutive AS1 and AS2, we can infer that there is a direct link between AS1 and AS2. Therefore, by looking at the flow of announced paths coming from neighboring ASes one can infer an AS level graph connecting ASes that have appeared at least on one of the announced AS paths \cite{Gao:2001:SIR:504611.504612}. We leverage this approach to create each minute a snapshot of the AS level graph. For this purpose we combine several BGP feeds coming from several BGP collectors. As announcements are made at prefix level, we can assign to each link in the AS level graph a {\tt count} attribute that measure the number of IP addresses that can be reached through paths going through the link. We moreover assign to each vertex in the graph some attributes including the number of prefixes issued at the AS represented by the node, and {\tt ctime} the last time there was an update or an announcement relative to that AS.


Indeed, the AS graphs generated by this approach are known to be incomplete \cite{Cheno}. In particular, BGP path filtering policies do not expose less preferred paths that would be chosen if the preferred announced path were not available. Moreover, traffic engineering and load balancing can further blur the real AS graph \cite{Roughan2011}. This incompleteness adds a dimension to the challenge addressed in this paper. The question we will address in this paper is "can we detect major changes in Internet structure by just looking at incomplete AS graphs that are gathered through a limited number of collection points".

\subsection{Riemanian Manifolds}
\label{sec:RiemanianManifolds}
Intuitively, a manifold is a surface in a multi-dimensional space that can be locally mapped to an euclidean space, {\em i.e.}, if we zoom in enough on the surface it resembles an euclidean plane. A Riemannian manifold is one that have attached to it a bilinear distance metric, which then allows for a notion of angles and distances on the manifold. One can look at a manifold from two perspectives: the extrinsic view considers the manifold as an object embedded in a (larger) euclidean space, while the intrinsic view considers properties determined solely by distance within the surface. In other terms, the extrinsic view is the one of an observer that could look at the manifold globally, and the intrinsic view is that of an ant (or an army of ants) moving along the manifold.
Relation between these two perspectives has been a major question in geometry, \emph{e.g.} the "remarkable theorem" of Gauss, {\em Theorema Egregium}, states that the curvature of a surface does not change if one bends the surface without stretching it. Thus the curvature is an intrinsic invariant of a surface \cite{Berger}. The intrinsic and extrinsic viewpoints are fundamental in our work as well, since we aim at characterizing which local changes will result in major global changes. Through the {\em Theorema Egregium} the notion of curvature of a surface becomes a main tool for doing this. 

\begin{figure}[ht]%
    \centering
    \includegraphics[width=8cm]{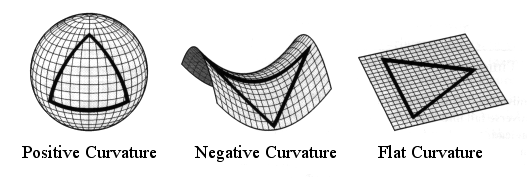} \caption{In flat geometry, parallel lines are equidistant while they tend to get further ({\em resp.} closer) in hyperbolic ({\em  resp.} spherical) space.} %
    \label{fig:curvature}%
\end{figure}

Curvature can be intimately related to the local behavior of geodesics, {\em i.e.}, shortest paths along the manifold. On a space with (strictly) positive curvature, two distinct geodesics starting at two points close to each other and pointing to the same destination will ultimately converge to the same point \cite{Ollivier}. Inversely  with negative curvature, two geodesics will drift apart getting further of each other (see Figure~\ref{fig:curvature}).
Several metrics exist for describing the curvature such as the sectional curvature \cite{Berger}, Gromov $\delta$-hyperbolicity \cite{Chen}, Bakry {\'E}mery Tensor \cite{Bakry}, Forman-Ricci curvature \cite{Forman2003} \cite{Weber} {\em etc.} \cite{Berger2003}. 
In this paper, we use an extension of the Ricci curvature to discrete objects, and particularly graphs, which we describe in the following section.

\subsection{Discrete curvature and the Ollivier-Ricci curvature}
\label{sec:ricci}

Let's define the discrete extension of Ricci curvature to a graph $G = (V,E)$. This extension uses the concept of transportation cost $d$ \cite{Piccoli} over a graph. Let's assume a graph where a distance metric $d(x,y)$ is defined between each pair of vertices $(x,y) \in V \times V$. 
This distance can be the minimum hop distance, a minimal weight distance or any other arbitrary distance matrix. 
Consider distributions $\mu$ and $\nu$ over the vertex set $V$, i.e. $\mu(x) \geq 0$  for $x \in V$ and $\sum_{x \in V} \mu(x) = 1$.
The optimal transport $\theta^*(\mu,\nu)$ is defined as follows.
\begin{definition}[Optimal Transport]
\begin{align}
\gamma^*(\mu,\nu) = \arg\min_{\theta} \sum_{x,y \in V} \gamma(x,y) d(x,y) \label{eq:optTransport},
\end{align}
where the minimization is over all transport plans $\theta$, \emph{i.e.} $\theta : V\times V \in \mathbb{R}$ such that
\begin{align}
\theta(x,y) \geq 0 \quad &\text{ for all } x,y \in V \nonumber \\
\sum_{y \in V} \theta(x,y) = \mu(x) \quad &\text{ for all } x \in V \nonumber \\
\sum_{x \in V} \theta(x,y) = \nu(y) \quad &\text{ for all } y \in V.
\end{align}
The value $C(\theta^*,\mu,\nu) \triangleq \sum_{x,y \in V} \theta^*(\mu,\nu) d(x,y)$ is referred to as the {\em transportation distance}.
\end{definition}
Intuitively, the transportation distance evaluates the effort needed to transport a mass distributed following a distribution $\mu(x)$ over the different vertices of the graph to another mass distribution $\nu(x)$. The optimal transport $\theta^*(\mu,\nu)$ always exists and can be computed exactly by solving the linear program \eqref{eq:optTransport}, or approximated efficiently (see for example \cite{Cuturi}). 

We can leverage the optimal transport to extend the Ricci curvature to graphs.
\begin{definition}[Ollivier's Ricci curvature \cite{Ollivier:2009}]
The Ollivier Ricci curvature $\kappa(x,y)$ between two vertices $x$ and $y \in V$ is defined as:
\begin{equation}
    \kappa(x,y) = 1 - \frac{C(\theta^*, \mu_x,\mu_y)}{d(x,y)},
\end{equation}
where $\{ \mu_z$ for $z \in V \}$ is a family of distribution.
\end{definition}
The distributions $\mu_z$ can be general, and we will specify which distribution is suitable for our application later on.
It can be shown that the curvature is bounded in general as $-2 \le \kappa(x,y) \le 1$ \cite{Ollivier}. However for distribution $\mu_z$ such that half of the mass is positioned on the central node and the remaining mass is distributed over the neighbors, one can prove that $-1 \le \kappa(x,y) \le 1$ \cite{Shiping}.

\noindent \textbf{Examples:} To illustrate intuitively the Ollivier-Ricci curvature let us look at three extreme cases. For these examples, we assume $\mu_z$ put weights uniformly among neighbors of $z$, \emph{i.e.} if $d_z$ is the number of neighbors of vertex $z$, then $\mu_z(x) = \frac{1}{d_z}$ if $x$ is a neighbor of $z$, and 0 otherwise.
\begin{itemize}
\item Let $G = (V,E)$ be a clique with $N$ vertices. In this case all neighbors of $x$ are neighbors of $y$. So the optimal transport plan $\theta^*(\mu_x,\mu_y)$ has not to transport anything from neighbors of $x$ to neighbors of $y$ and only needs to transport a mass $\frac{1}{N-1}$ from $y$ to $x$. This means that the Ollivier-Ricci curvature equal to  $\kappa(x,y)=1-\frac{1}{N-1}$. As $N$ grows, the curvature tends to its maximal value $1$. 
\item Let $G$ now be an alignment of nodes, {\em i.e.}, a line graph, and let $x$ and $y$ be any two distinct nodes on the graph. It is easy to verify that the optimal transport $\theta^*(\mu_x,\mu_y)$ corresponds to transporting the mass in the left (resp. right) neighbor  of $x$ to the left (resp. right) neighbor of $y$ with a cost of $d(x,y)$, therefore $\kappa(x,y) = 0$.
\item Let $G$ be two $N$ nodes star networks connected by a single link. In this case, the optimal transport plan  between the first star center $x$ and the second center $y$ send all the masses in neighbors of $x$ to neighbors of $y$ through the link connecting the two stars with a distance cost of 3 and transport the mass in $y$ to $x$ with a distance cost of 1. This means that the optimal cost will be $3-\frac{2}{N}$  resulting into an Ollivier-Ricci curvature $\kappa(x,y) = -2+\frac{2}{N}$. It is noteworthy that if the distribution $\mu_x(z)$ is such that half of the masses are positioned on $x$ and $y$ respectively with the remaining masses are distributed evenly on their neighbors, 
the optimal transport cost becomes $2-\frac{3}{2N}$ and the Ollivier-Ricci curvature is $\kappa(x,y) = -1+\frac{3}{2N}$
\end{itemize}
Real graphs like the AS level graphs are not as simple as the above examples. Nonetheless, when the Ollivier-Ricci curvature obtained between two vertices is close to 1 this can be interpreted as the two nodes are in a clique-like structure, {\em i.e.}, there is a rich network of link connecting the two nodes such that almost all nodes are neighbors of each other. On the other hand, when we have negative curvature close to $-2$,
all the paths connecting the two vertices are fully relying on a few nodes. This interpretation will be helpful to understand the curvature variation, {\em i.e.}, increase of curvature means that the network is becoming more connected and decrease of curvature is a sign of the emergence of a bottleneck.

\section{Ollivier-Ricci Embedding Monitoring System}
\label{sec:Ollivier}
In this section we will describe how to implement the Ollivier-Ricci Embedding of the AS level graph. Recall that the embedding consists of projecting the AS level graph into a $K$-dimensional Euclidean subspace where each coordinate is the Ricci curvature measured from a node of the graph (an AS) to one of the $K$ landmarks. We will define the curvature matrix as the $N \times K$ matrix where each line contains the $K$ dimensional projection of an AS that have seen a BGP update during the graph collection snapshot. 

\subsection{Pre-processing steps}

The previous section briefly introduced some background of the theoretical tools used by the monitoring system we describe here. We introduced the Ollivier-Ricci curvature as a metric relating local property of a vertex to global properties of the graph. 

We assume access to a series of (periodic) snapshots of the AS-level BGP graphs. Each vertex representing an AS in this graph has a set of attributes including a timestamp {\tt ctime} identifying the last time the AS has seen a prefix update or withdrawal announcement, as well as the number of prefixes it announces. Each link also contains a {\tt count} attribute measuring the number of prefixes that are announced to be reachable through this link. 

\begin{figure*}[ht]%
    \centering
    \subfloat[Global effect event]{{\includegraphics[width=5.5
    cm]{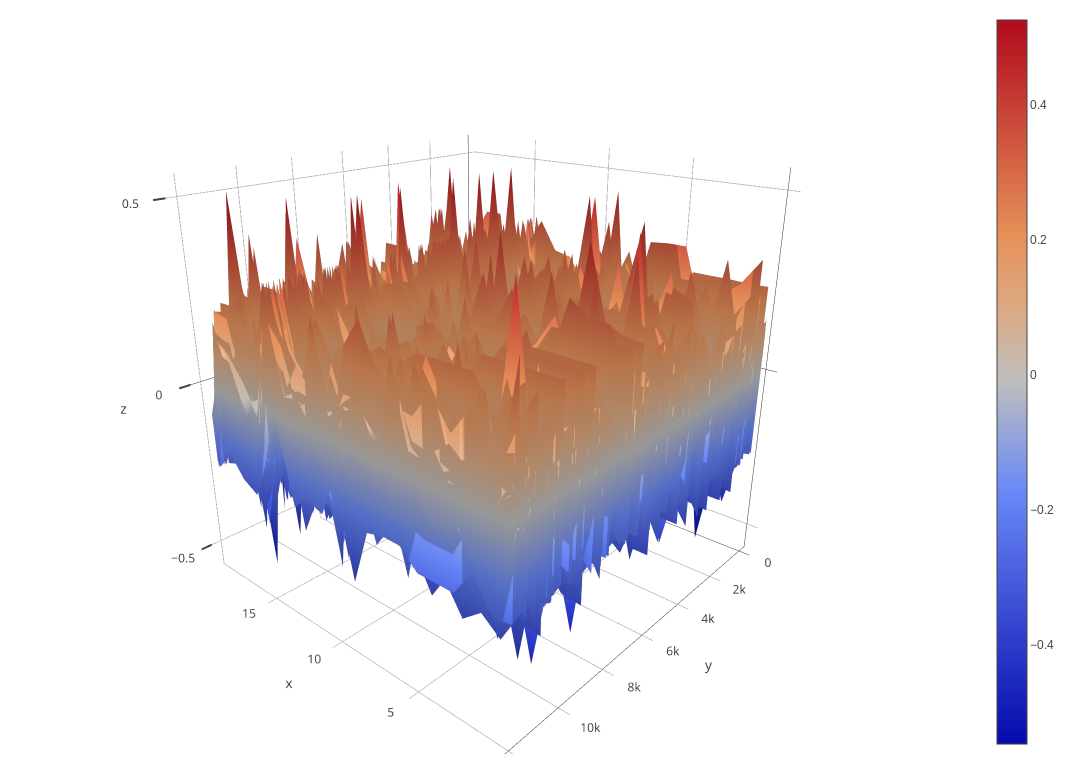}}\label{fig:global}}
    \subfloat[Local effect event]{{\includegraphics[width=5.5
    cm]{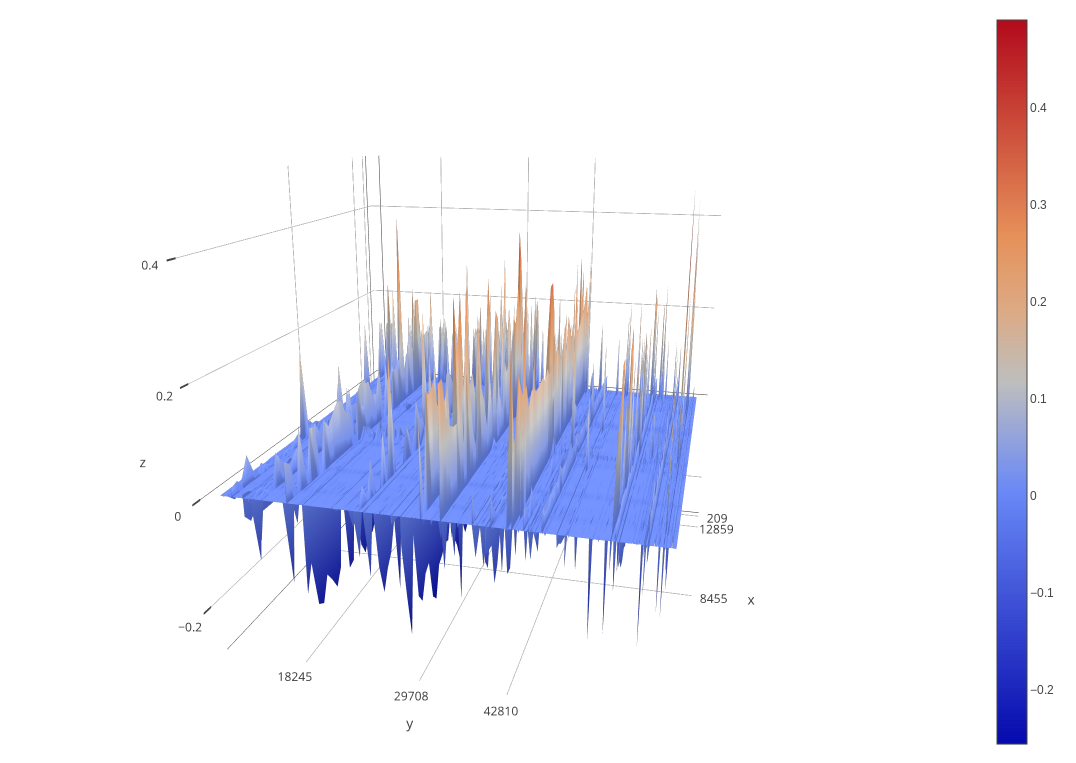}}\label{fig:local}}
    \subfloat[Natural drift]{{\includegraphics[width=5.5cm]{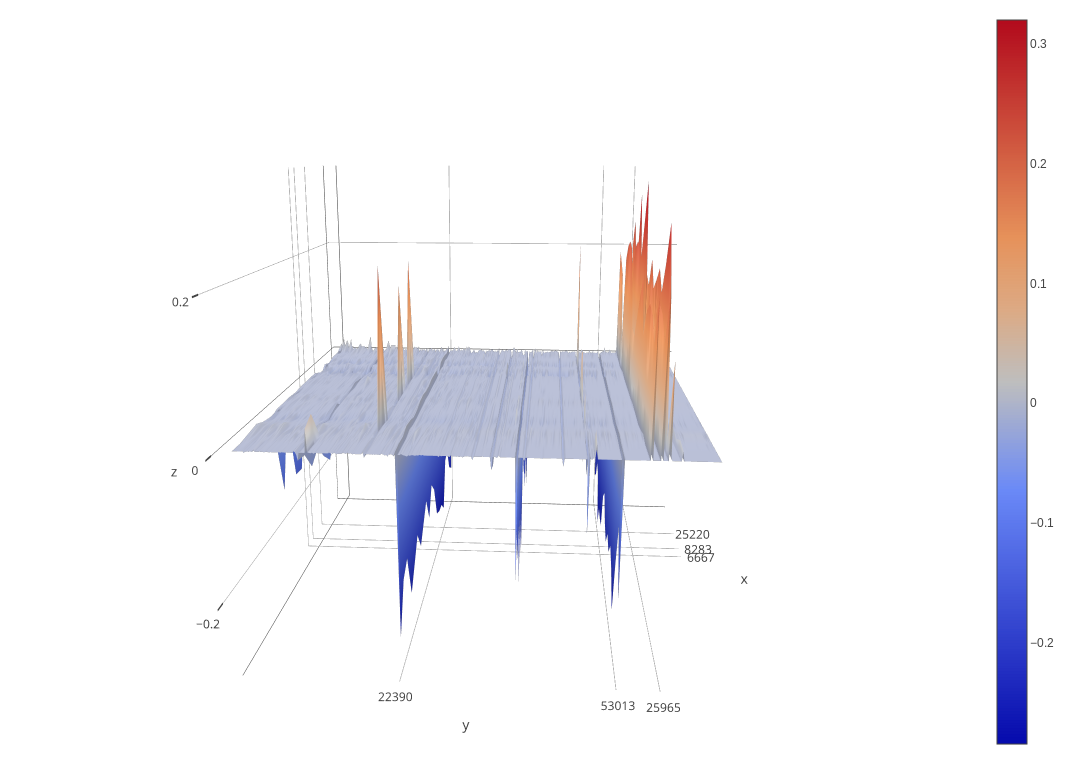} }    \label{fig:drift}}%
    \qquad
    \hspace{12cm}
    \subfloat{\hbox{\hspace{0.5em}{\includegraphics[width=5.5cm,height=4.2cm]{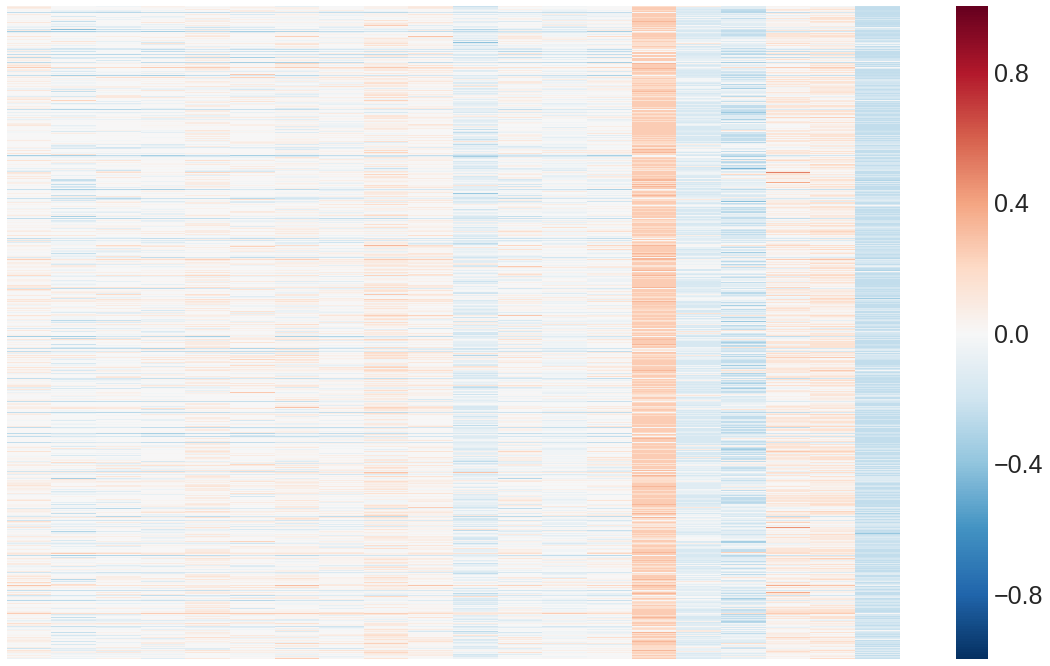}}}}
    \subfloat{{\includegraphics[width=5.5
    cm]{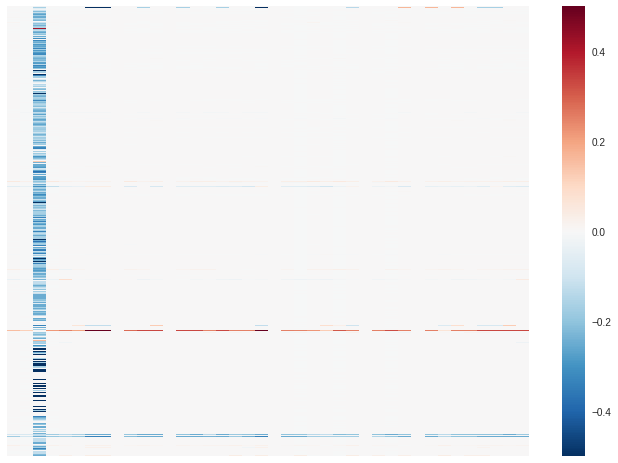}}}%
    \subfloat{{\includegraphics[width=5.5cm]{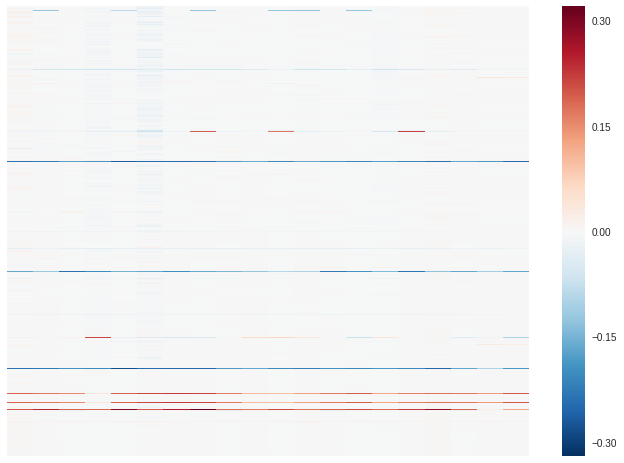}}\label{fig:example}}
    \caption{Representative sample of the matrix $\Delta^k$ for 3 different types of events. In Fig. \ref{fig:global} we show the Google leak Incident that impacted Japan the 25th of August 2017. In Fig.\ref{fig:local} we show a local BGP event, an outage, that impacts only slightly a small part of the matrix (Brazil leak, see table \ref{tab:majorevents}). The event pictured  in Fig. \ref{fig:drift} corresponds to a period (5 Jan 2018, see in table \ref{tab:majorevents}) where there is no either local, outage or hijack, or global event in the network. However, there will still be a natural drift coming from BGP updates and withdrawals.}.    \label{fig:3D}
\end{figure*}

\noindent \textbf{Choice of mass distribution:}

In order to derive the curvature we have to set a source and destination mass distribution for the transport ($\mu_x$ and $\mu_y$ in \eqref{eq:optTransport}). In his seminal paper, Ollivier \cite{Ollivier:2009} used a simplistic distribution that distributes evenly the mass to transport over the neighbors of the vertex.  As explained in Section \ref{sec:ricci}, we will put a mass in the center node in order to ensure that the Ollivier-Ricci curvature between each two nodes $x$ and $y$ remains in $-1\le \kappa(x,y) \le 1$. 

However, this distribution assumes the same importance for all neighbors which might be unrealistic. A good choice for this distribution is to set it according to the amount of traffic flowing from one AS to another. This information is often unavailable for large networks. Alternatively, we need to use a measure that can be readily obtained from BGP feeds.  

As we store for each AS to AS link in the network a {\tt count} attribute that contains the number of IP addresses reachable through this link, we set the source and destination mass distributions following the distribution of {\tt count} to the neighbors. This distribution, that is recalculated after each update or withdrawal, gives to each neighbor a weight proportional to the number of IP addresses that is accessible through them. Next, we use the Ollivier-Ricci curvature derived from this measure.

\noindent \textbf{Construction of curvature matrix:}

In its steady state, the AS graph contains up to 80 K ASes and 200 K  links that dynamically change because of permanent updates and withdrawals. It is therefore unfeasible to monitor the variation of curvature between all ASes and landmarks. We therefore resort to use the {\tt ctime} attribute to detect ASes that have undergone a change and limit the computation to those ASes.

Let's assume that during the snapshot $k$, $m$ ASes have seen a prefix update or withdrawal. We derive for the snapshot $k$ a $m\times L$ matrix $\mathcal{C}_k$, where $L$ is the number of landmarks. 

In order to detect changes we generate a similar matrix of curvatures with the same $m\times L$ size, but for the last snapshot $k-1$. We then use $\Delta^k=\mathcal{C}_k-\mathcal{C}_{k-1}$ as the basic statistics for changes detection. We show three examples of the $\Delta$ matrix corresponding to three different BGP events in Figure \ref{fig:3D}.

\subsection{Monitoring}
\label{sec:embedding}

A dynamic graph will undergo continuous changes that will alter the graph locally and potentially lead to small modifications at the global level. This results in permanent fluctuations of the curvature measure. Nonetheless, the Ollivier-Ricci variations can become large when an event changes the geometry and impacts the geodesics. Every non-zero value in the $\mathcal{C}_k$, corresponds to either a stretch or a compression of the underlying geometry and indicates a change in the network conditions, possibly induced by an anomaly. 

The sign of the spikes characterizes the effect the event had on the connectivity. A positive spike means that the curvature has increased, and that the neighborhoods of the changed AS are now more connected to the rest of the network. Inversely, a negative spike implies that the neighborhoods of the changed AS are more dependent on a reduced number of connections with the landmarks. Generally, a BGP hijacks or leaks generate a positive shift in the Ricci curvature as they involve for the impacted vertex the emergence of new and shorter paths \cite{Survey-bis}. Inversely, link (route) failures generally decrease the curvature and the diversity (see for example Fig. \ref{fig:local}).

\noindent \textbf{Interpretation of Curvature Matrix} 

As can be seen in Fig. \ref{fig:3D}, the matrix $\Delta$ contains vertical and horizontal alignments of decreases or increases in curvature.  A column with large variations of curvature means that the geometry toward the landmark attached to this column has experienced a major change. If several landmarks experience important changes, it indicates a global change. Similarly, a horizontal alignment of large variations means that the AS has seen major change toward all landmarks. A global change will induce several horizontal and vertical lines with important changes. We see also in Fig. \ref{fig:drift} a representative case showing the perpetual drift. Generally, the drift will show itself with changes in the curvature that are limited to the impacted ASes (represented by horizontal lines in the heatmap of the curvature matrix). Local events and global events will exhibit vertical lines in the heatmap, meaning that at least one of the landmarks has seen curvature change to numerous ASes. The difference between a local and a global event is in the strength of the change and the number of large curvature changes. While the previous analysis is intuitive, we need to define an automatized detection scheme.

\begin{figure}[h]
\begin{center}
  \includegraphics[width=8 cm, height=4cm] {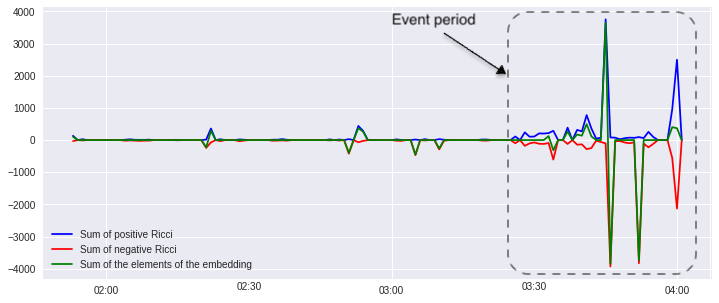}
  \caption{Time series of the sum of the positive (resp. negative) elements in $\Delta_{k}$ along with the global sum.}
\label{fig:diff}
\end{center}
\end{figure}

We show in Fig \ref{fig:diff} the sum of positive and negative values in the $\Delta^k$ matrix. Following the discussion in section \ref{sec:embedding}, we observe a behavior compatible with the Gauss-Bonnet theorem, {\em i.e.} zero sum or very small one in period without major BGP events, and a change in period with large event.

From the previous discussion one can consider that monitoring the sum of curvature is enough to detect large events. Nonetheless, when the Gauss-Bonnet theorem 
is valid, the sum of curvature will not change whatever large is the fluctuation. However, the Gauss-Bonnet theorem is not formally valid for Ollivier-Ricci curvature. This means, just looking at the sum of curvature might generate false alarms and misdetections. We have therefore to define another method to translate the matrix $\Delta^k$ into a time-series over which we will detect large events.

\noindent \textbf{Analysis of Curvature Matrix}As said before columns or lines structure in the matrix $m\times L$ matrix $\Delta^k$ are representative of the changes in the topology. Typically, the matrix contains several very small values, relative to small curvature changes, and a limited number of lines or columns with larger curvature changes. In order to analyze the structure of this matrix we will use its singular values \footnote{Singular values are extension of eigenvalues to non-square matrices \cite{singular}}, {\em i.e.}, the roots of eigenvalues of the $L\times L$ matrix ${\Delta^k}^T\Delta^k$.  The elements on the diagonal of ${\Delta^k}^T\Delta^k$ are the column-wise sum of squares of $\Delta^k$ matrix and its trace is the Frobenius norm of $\Delta^k$, $\|\Delta^k\|_F$ :
$$
\|\Delta^k\|_F=\sum_i \sum_j \left(\delta^k_{ij}\right)^2.
$$
The Frobenius norm represent therefore the variance of the values in $\Delta^k$, {\em i.e.}, the strength of the curvature changes.

One can check that if the matrix $\Delta^k$ contains a $l$ non-zero column, {\em i.e.}, its rank is $l$, the  matrix ${\Delta^k}^T\Delta^k$  will have $l$ strictly positive eigenvalues, or equivalently $\Delta^k$ will have $l$ non-zero singular values. However, in practice, the curvature matrix is filled with small values, rather than 0, and only some of the lines or columns will have large values. In such case, $\Delta^k$ have some large singular values  ${\Delta^k}^T\Delta^k$ representing the large values columns, and other eigenvalues close to 0. In such contexts, the stable rank defined as
$$
\gamma_k=\frac{\|\Delta^k\|_F}{\lambda^0_k},
$$
where $\lambda^0_k$ is the largest eigenvalue of ${\Delta^k}^T\Delta^k$. $\gamma_k$ is frequently used as a robust estimator of the rank of a matrix, in particular for generating low rank approximation of noisy matrices \cite{stablerank}. As the rank is directly related to number of large columns in $\Delta^k$, the stable rank estimates this number, {\em i.e.}, a stable rank close to 1 means that only a single column of $\Delta^k$ has important changes, while a large stable rank indicates large number of columns, {\em i.e.}, landmarks, having seen important changes. It is also noteworthy that $\lambda^0_k$ have the below property :
$$
\lambda^0_k=\max_{\|X\|_2 \neq 0}\frac{\|\Delta^k X\|_2}{\|X\|_2}.
$$
In other terms, the largest singular value gives the largest norm stretching that the $\Delta^k$ matrix can generate.

The above observations lead into a simple anomaly detection scheme. For each incoming matrix $\Delta$, we evaluate the normalized Frobenius norm, {\em i.e.}, $\overline{\|\Delta^k\|_F}=\frac{\|\Delta^k\|_F}{m}$, where $m$ is the number of ASes having seen changes in the $k^\textrm{th}$ snapshot, and the stable rank of $\Delta^k$, {\em i.e.}, $\gamma_k$. The set $\left \{ \left(\bar{\|\Delta^k\|_F}, \gamma^{-1}_k \right), \; k=0,\ldots, \right\}$ defines a time trajectory that will be used to detect large events. We use $\gamma^{-1}_k$ in place of $\gamma^{-1}_k$, as it is bounded, {\em i.e.}, $0<\gamma^{-1}_k<1$. A large event will have a large value of  $\overline{\|\Delta^k\|_F}$ and a small $\gamma_k^{-1}$, while a local event might have large value of $\overline{\|\Delta^k\|_F}$  but will have a value of $\gamma_k$ close to 1.

\begin{figure*}[ht]
\begin{center}
  \includegraphics[width=\textwidth]{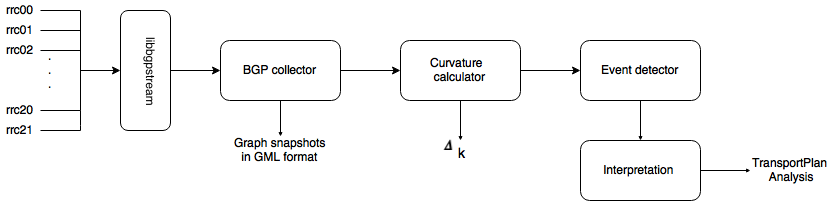}
  \caption{Components of the monitoring system}
  \label{fig:monitoringsys}
\end{center}
\end{figure*}

\subsection{Landmarks selection}
\label{sec:landmarks}
Our approach relies on constructing the Ollivier-Ricci embedding in $\mathbb{R}^N$. However, monitoring the variations of all distances is unfeasible over a real AS-level graph with over 60k vertices and 3.6 billion distances \cite{cidr}. 
We reduce the dimensionality by choosing a set of $L<<N$ landmarks and only monitor the variation of curvature towards them, {\em i.e.}, we represent the position of vertex $i$ by a vector ${\bf X}^i=(x^i_j),\; j=1, \ldots L$ where $x^i_j=\kappa(i,j)$ is the curvature from $i$ to landmark $j$.
The essential property we are seeking to preserve despite the dimensionality reduction, is that the variation of the time series between time $t$ and $t+1$, denoted hereafter by the {\em drift}, should be small when there are no major changes in the graph underlying geometry and substantial when such a large change happens. The landmarks should satisfy this property and we will validate this claim in Sec.~\ref{sec:evaluation}. 

Choosing landmarks in a highly dynamic graph is a complex task \cite{Leskovec}. 

We considered multiple landmark selection process: (a) random landmarks selection, (b) highest degree, (c) highest centrality, (d) highest number of triangles, (e) Tier 1 and Tier 2 AS, a random walk starting at (e) a random vertex or (g) at the collector node, and finally (f) a lazy random walk that we introduced and which was developed for this purpose.

In order to evaluate the performance of different landmark selection methods, we define two metrics that are calculated over a landmark set $R$. A first metric, $S_1$, evaluates the diversity arising from $R$, by assessing the proportion of distinct neighbors of vertices in $R$ among all possible neighbors, and $S_2$ the average distance in hops between landmarks, {\em i.e.}:
\begin{equation}
\label{Eq:S1}
\begin{split}
	 S_1(R) &= \frac{\left| \bigcup\limits_{v \in R} N(v)\right|}{\sum_{w \in R}\left| N(w)\right|}\\
 	S_2(R) &= \frac{1}{2|R|}\sum_{v \in R} \sum_{w \in R} d(v,w)
\end{split}
\end{equation}
High diversity is a necessary condition for a landmark to have high curvature. So we use the $S_1$ metric as a low complexity proxy to detecting potentially high curvatures.  A good set of landmark should therefore at the same time achieve a good diversity, large value of $S_1$, and be far apart, large value of $S_2$.
We consider 30 of our collected AS graphs and run over them the different landmark selection algorithm and see the outcomes. For random walk based approaches (lazy random walk, fully random and random walk), we do 10 runs each with 35k random steps (10k for 10 landmarks) and for deterministic approach (triangle, degree, tier 1\& 2), we execute the code over the 30 graphs. We compute for each case the score $S_1$ and $S_2$ of resulting landmarks obtained by different methods. We report the average of $S_2$ in \ref{table:1} and look at the distribution of the measure $S_1$ for 20 landmarks in \ref{fig:comp}.
\begin{figure}[ht]
 {\includegraphics[width=9cm]{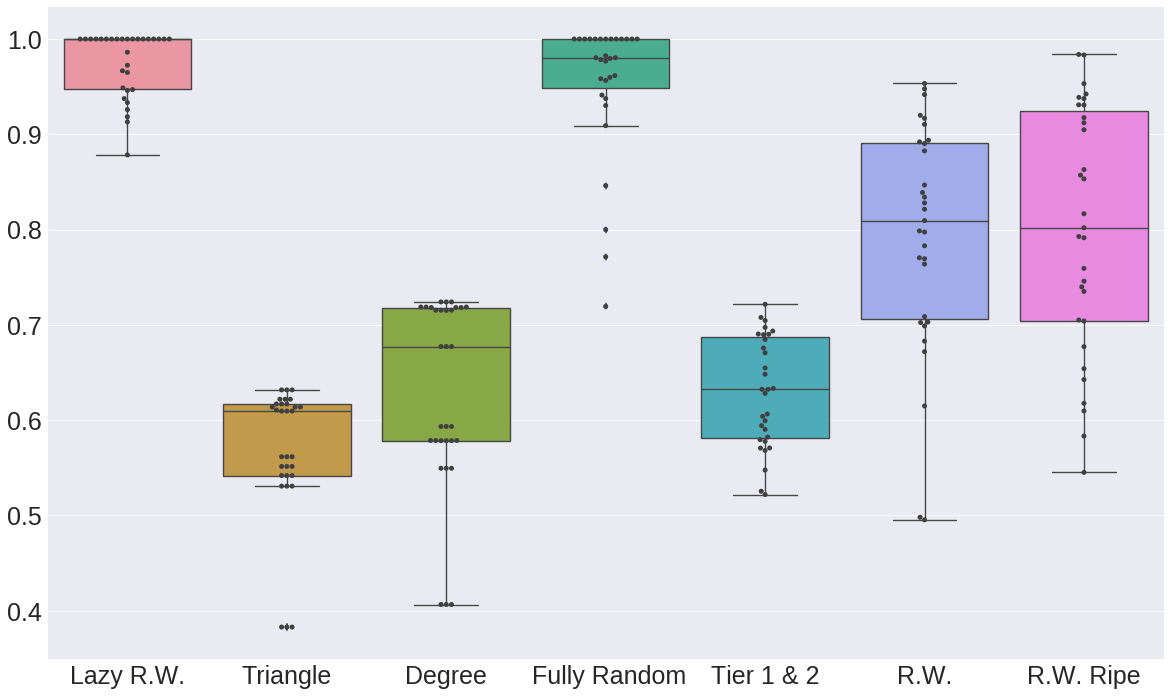}\label{fig:comp}}
\caption{Distribution of $S_1$ for 20 landmarks for different landmarks selection approaches}
\label{fig:comp}
\end{figure}
\begin{table}[h!]
\centering
 \begin{tabular}{ |p{2.7 cm}||p{1.5cm}|p{1.5cm}|p{1.5cm}|  }
 \hline
 \multicolumn{4}{|c|}{Landmarks} \\
 \hline
 $\#$ of landmarks & $\#$10 (10K) & $\#$ 20 (35K) & $\#$ 30 (35k) \\
 \hline
 Lazy Rand Walk & \textbf{5.447742}   & \textbf{7.07428}  &\textbf{5.2616129}\\
 Fully random   & 3.670968   &4.118387&   3.841935\\
 Highest degree &   1.123871  & 1.265484  & 1.106452\\
 Highest centrality & 1.16788 & 1.278206 & 1.125779 \\
 Highest triangle    & 1.144516  & 1.209032 &  0.984516\\
 Mix Tier 1 and Tier 2 & 1.565161   & 1.548065 &  1.620645\\
 Rand. Walk & 1.859355 & 2.0933387 &1.830323\\
 Rand. Walk Ripe & 1.741935 & 1.720806 &1.823971\\
 \hline
\end{tabular}
\caption{Average of $S_2$ for different landmarks selection approaches}
\label{table:1}
\end{table}
We can see that the lazy random walk achieves a better performance both in term of diversity and of average distance, we will therefore use the 20 landmarks coming from this method in the forthcoming.


{
}

\

\section{Evaluation and validation}
\label{sec:evaluation}
The previous sections presented several components of a large-scale graph monitoring system and its application to the particular case of AS level BGP graph. We present in Fig. \ref{fig:monitoringsys} the components of the monitoring system. In this section, we will assert the performance of our approach and compare it with other approaches.
\subsection{Datasets}
The first stage of the monitoring system pipeline is the BGP feeds gathering. 
For the results shown in this paper we used the python version of libBGPstream, pybgpstream \cite{libbgpstream} and gathered BGP feeds coming from all 13 active RIPE Routing Information Service open feeds \cite{RIPE}. The BGP gathering tools use the approach described in section \ref{sec:BGP} to update an AS level graph that is saved each minute into graph snapshot in a GML formatted file. Each trace is initiated with an empty graph, meaning that there is a transient phase during which the AS level graph size increases according to arriving BGP updates. Generally after up to one hour, the graph reaches a steady state with around 68 k ASes. It is noteworthy that even in steady state the graph is permanently updated with new paths that appear and one can never assume that the AS level graph is becoming static. Because of the permanent dynamic of the AS level graph, we are making the analysis over all snapshots regardless of being in transient phase or steady state. As we will see this can result in observing relatively large geometry changes that are arising from gaining visibility for a large part of the topology. 

As historical data are available going back for example to 1999 for the {\tt rrc00}, we have reconstructed the BGP feeds as they were announced during 8 different major BGP even of the past years (see table \ref{tab:majorevents} for a list of monitored events). In addition to these events, we also run the BGP gathering components over long period without known major events. 

\begin{table}[ht]
    \centering
    \begin{tabular}{|p{1cm}|p{1.5cm}|p{1cm}|p{1cm}|p{1cm}|p{1cm}|}
 	\hline
 Date & Name & Trace duration &  \#ASes& \#links & Extend of the event\\
 \hline
  23 Jan 2008    & Middle East cable cut  & 7h00  & $\approx 22 K$ &  $\approx 40$ K  & Medium \\
  \hline
  25 Aug 2017 & Google leakage   & 3h30  & $\approx 60$K & $\approx 120$K & Large\\
 \hline
 21 Oct 2017 & Brazil leakage & 3h30 & $\approx 60$K & $\approx 100$K & Large\\
 \hline
  22 Apr 2016  &   Innofield incident & 1h10 & $\approx 8$K  & $\approx 20$K & Large \\
  \hline
  12 Dec 2017 &  Russia hijack & 2h00 & $\approx 12$K &$\approx 22$K & Large \\
  \hline
  21 Oct 2017 & Before Brazil leakage & 2h00 & $\approx 13$K & $\approx 35$K & No event \\
  \hline
  5 Jan 2018 & NA & 20h00 & $\approx 60$K & $\approx 100$K & No event\\
  \hline
  23 Apr 2016 & NA & 1h45 & $\approx 18$K & $\approx 40$K & Medium\\
  \hline
  22 Apr 2016 & After Innofield incident & 1h45 & $\approx 14$K & $\approx 30$K  & No event\\
  \hline
  21 Oct 2017 & Before Brazil Leakage & 8h30 & $\approx 55$K & $\approx 91$K & No event\\
  \hline
 \hline
    \end{tabular}
    \caption{List of major BGP events studied in this paper}
    \label{tab:majorevents}
\end{table}
\subsection{Complexity}
The different steps of the monitoring system have different complexities.  We are currently executing the BGP gathering step on a single computer using a multithreaded (4 threads) python code. We are observing over different trace an average of 1200 BGP announcements per second, that can peak to up to 10 000 during period of high intensity. The current non-optimized python code is running in almost real-time, {\em i.e.}, one second in the BGP feed is processed in less than one sec, beyond when BGP updates peak times when the current system lags. The memory usage of the tool is relatively large, in particular the routes database might contain up to 12 millions BGP route occupies around 6 GBytes of memory. 

The most costly step of the monitoring system is the derivation of the $\Delta$ matrix. Let's remind that this matrix contains $m\times L$ curvature elements. As explained before, we have used throughout the paper $L=20$ landmarks. The number of ASes, $m$, that have seen an update or withdrawal during  an AS graph snapshot interval varies accordingly to the dynamic of the network from 20 to more than 10000. This means that we need to solve up to 200000 linear programming optimization in the form given in Eq. \ref{eq:optTransport}. The main issue is that each one of these optimizations needs a distance matrix between all neighbors of source and destination vertices. Even if the calculating the shortest distance have a linear complexity calculating this matrix 200000 times over a graph with up to 60K nodes is challenging. Currently, using a single processor and  with a python code without major optimization, we are able to derive the $\Delta$ matrix and the event detection step, in average, in 3 minutes for each snapshot, {\em i.e.}, we have a 3 minutes of detection delay. We are confident that this time can be reduced with a better implementation in a more suitable language and over a multiprocessor machine. 

\subsection{Large-scale event detection validation} \label{sec:validation}
In this section we will see how the proposed monitoring system behaves in real operational settings. However, before dealing with our proposed system we have to compare it with alternative monitoring systems. Most of BGP monitoring systems target detection of localized event like outage and more importantly BGP hijacks. While these issues are indeed important, they generally have not a global impact. In \cite{Wang:2002} and \cite{MengChen2015} the BGP update volumes were used to detect and characterize large-scale BGP events. Both these papers showed that although the update volume is correlated to the scale of BGP events, this was not enough to construct a reliable detector, because of it is hard to predict and detect BGP sessions resets that can also result in large volume of updates. 

Other alternative embedding metrics that could be used in order to monitor changes toward landmarks are the shortest path distances and the spectral distances. The spectral distances are obtained through a spectral projection of the AS level graph \cite{Luxburg, monitoring2017}. We compare in Fig. \ref{fig:example} the distribution of the variations of these embedding metrics toward some landmarks and compare it with the Ollivier-Ricci embedding we have developed in this paper. We would like embedding metrics to exhibit small variations when strictly local events happen and large variations for large one. This means that the most of the time a suitable embedding metric should be around 0, with large value being less frequent. However, both shortest path distance and spectral distances suffer from high sensitivity to local changes. Any local change on the shortest distance might result at least to +1 or -1 on the embedding metric. Spectral distance is also know to be very sensitive to local neighborhood change, {\em i.e.}, node with same neighborhood are projected to the same spectral coordinates, and slight change in neighborhood can result in large variation of the spectral distance. The sensitivity of the two alternative metrics can be seen in Fig. \ref{fig:example}. The shortest distance metric show large discrete variation, and the spectral distance exhibit also frequently large variation, while the Ollivier-RRicci embedding show a higher concentration of values around 0 drift with less frequent large values. This indicates that the two alternative embedding metrics are likely to not have good performances for graph monitoring applications. We will show in the forthcoming that despite the two others Ollivier-Ricci embedding attain very good performance for monitoring large-scale graphs.
\begin{figure*}[h!]%
    \centering
    \subfloat[Shortest path embedding ]{{\includegraphics[width=6
    cm]{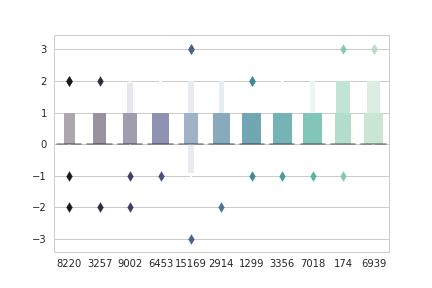} }}
    \subfloat[Spectral embedding ]{{\includegraphics[width=6
    cm]{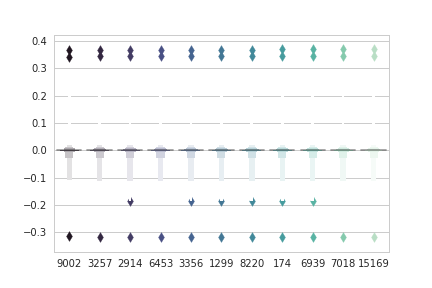}}}
    \subfloat[Ricci embedding]{{\includegraphics[width=6cm]{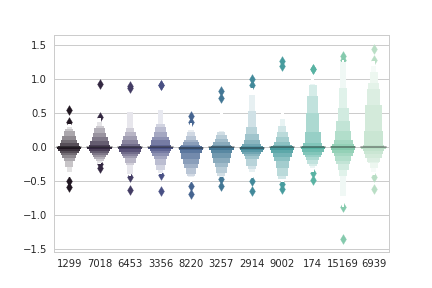} }}
    \label{fig:ricciVar}%
	\caption{Distribution of the variations of 3 possible embedding metrics}
    \label{fig:example}%
\end{figure*}

\begin{figure*}[ht]%
    \centering
    \subfloat[Google Leakage]{{\includegraphics[width=6.5cm] {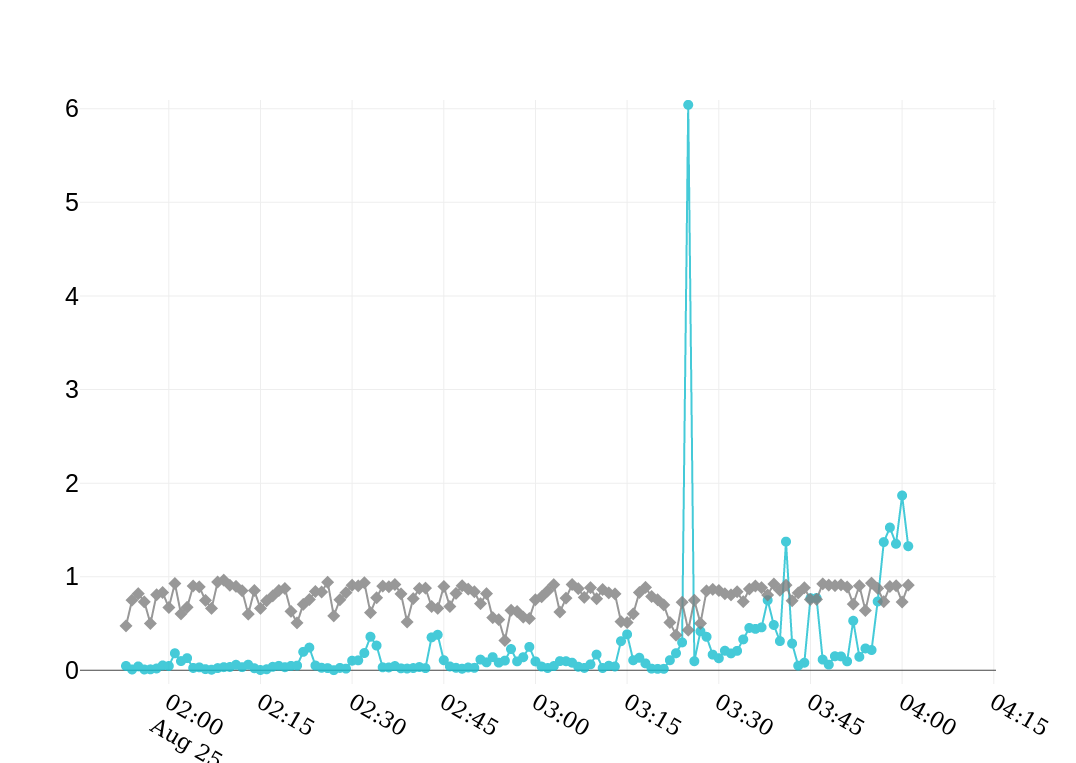}} \label{fig:phase1}}
    \subfloat[Phase diagram of the Google leakage]{{\includegraphics[width=5 cm]{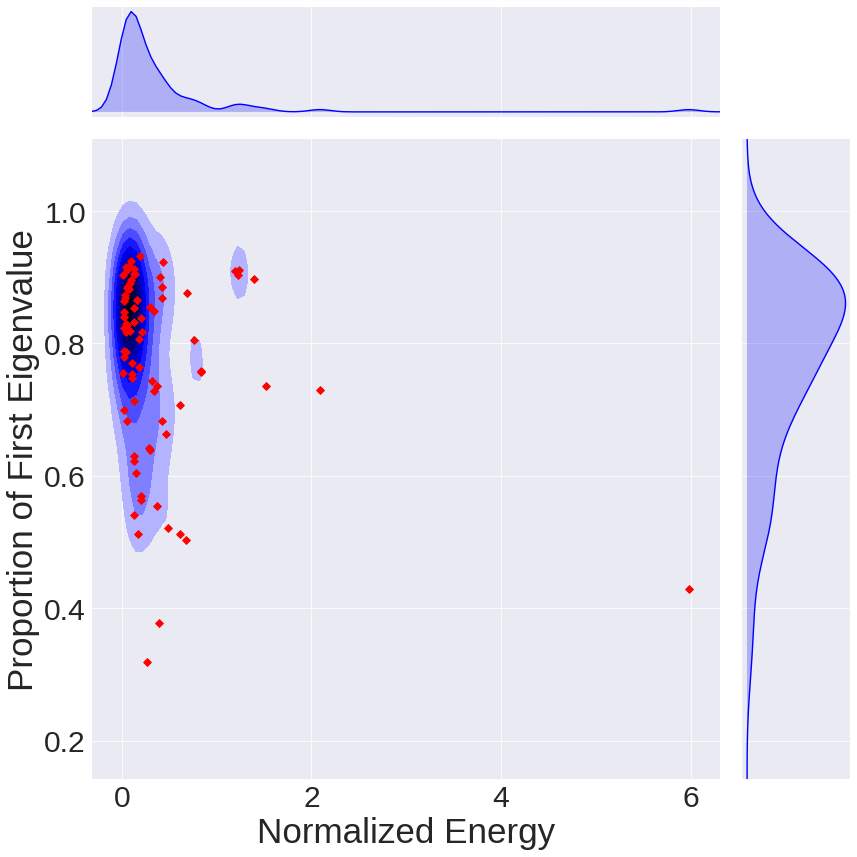}}\label{fig:phase2}}
    \subfloat[Brazil Leakage]{{\includegraphics[width=6.5
    cm]{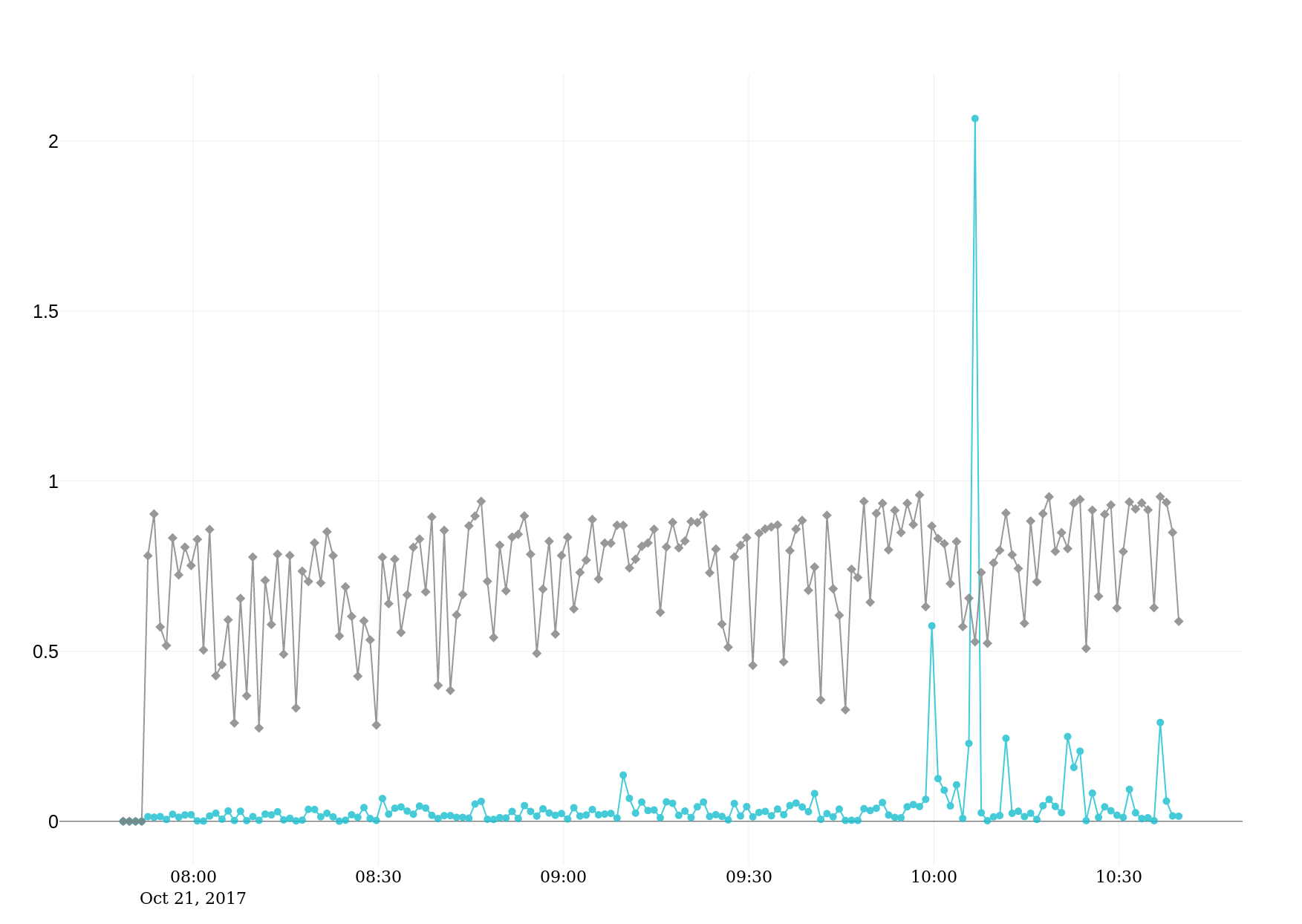} }}
    \caption{Validation of the large-scale events detection}
    \label{fig:Phase}
\end{figure*}

We illustrate the method for large-xfscale even detection by showing in Fig. \ref{fig:phase1} the time evolution of $\|\Delta^k\|_F$ obtained over AS level snapshots gathered around  Aug. 25, 2017. As can be seen from this diagram at 03:23 GMT the normalized energy $\overline{\|\Delta^k\|_F}$  exhibits a large peak. At 03:22 GMT, it has been reported that in Chicago, Illinois, Google generated a massive BGP leaks, that had major consequences for the Internet in Japan, and therefore the peak observed in \ref{fig:phase1} is related to this leak. The peak is well separated from other points of the phase diagram showing that Google BGP leak changed severely overall Internet geometry. A more in-depth  analysis of the $\Delta$ matrix shows a large increase of curvature, {\em i.e.}, the network becomes more similar to a clique and the overall shortest distances decrease. This is characteristic of a BGP leak that provide a lot of new paths to the network.  We show in Fig. \ref{fig:phase2}, the phase diagram, a 2 dimensional diagram with in the $x$-axis the normalized energy, $\overline{\|\Delta^k\|_F}$, and on the y-axis the inverse of the stable rank, $\gamma_k^{-1}$  relative to the previous time plot.As can be seen from phase diagram there are several points that all happen after 3:23 GMT with large $\overline{\|\Delta^k\|_F}>1$. All these points have overall decreasing curvature, {\em i.e.}, the new shortest paths leaked by the Google leakage are withdrawn gradually. However, for all these point $\gamma^{-1}_k > 0.7$, meaning that only a small number of landmarks are affected by the change, making the change local. In other term the global event at 03:22 GMT is resolved by a sequence of more limited local events. 

Similar behavior is observed for other large-scale event. We show in Fig. \ref{fig:phase3} another event that happened in Oct. 21, 2017, when a major BGP leakage happened in Brazil at 10:09 GMT. Here also we see a large peak of $\overline{\|\Delta^k\|_F}$ with $\gamma_k^{-1}<0.7$. 

\begin{figure}[h]
\centering
    \includegraphics[width=\columnwidth]{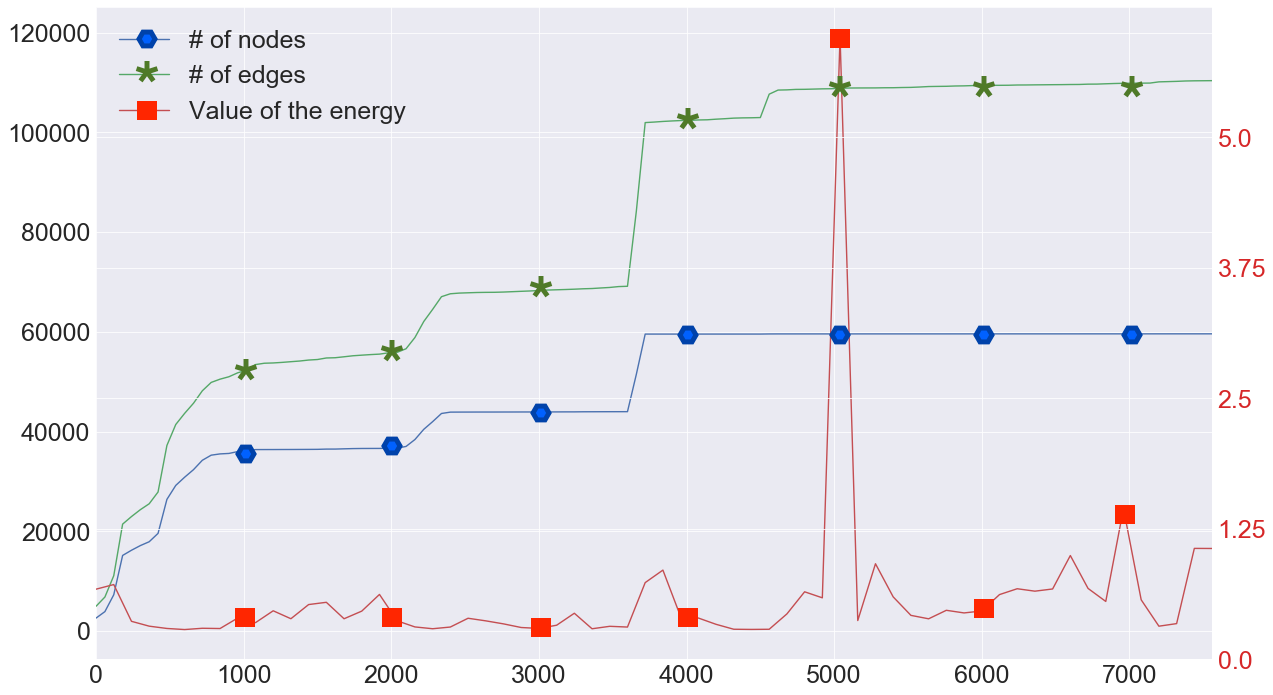}
    \caption{Time evolution the AS level graph AS nodes and links number}
  	\label{fig:timeEvolution}
\end{figure}

One typical question that arise is related to the dynamic building of the AS level graph, {\em i.e.} new ASes and link are appearing progressively in the graph. One can wonder if the addition of new nodes is not perturbing the detection process, resulting into spurious events that are caused by the collection process rather than real events over Internet. We show in Fig. \ref{fig:timeEvolution} the time evolution of the number of AS nodes and links in the AS graph along with the evolution of the energy in $\Delta^k$ matrix around a detected large-scale event. As can be seen from the picture at time of large increase in the number of nodes and links there is a slight increase in the  $\overline{\|\Delta^k\|_F}$, that is normal and expected, as overall by adding new nodes and links we are changing the structure of the network. However, there is no co-occurrence of large changes in with change in the number of nodes or links, meaning that large-scale events are not impacted by the dynamic process of building the AS graph on the fly. This is a remarkable property that show the robustness of the large-scale event detector.

\subsection{Analysis in the wild}
\label{sec:newevent}

\begin{figure*}[h!]%
    \centering
    \label{fig:checking}%
        \subfloat[Phase diagram of all datasets]{{\includegraphics[width=7 
    cm]{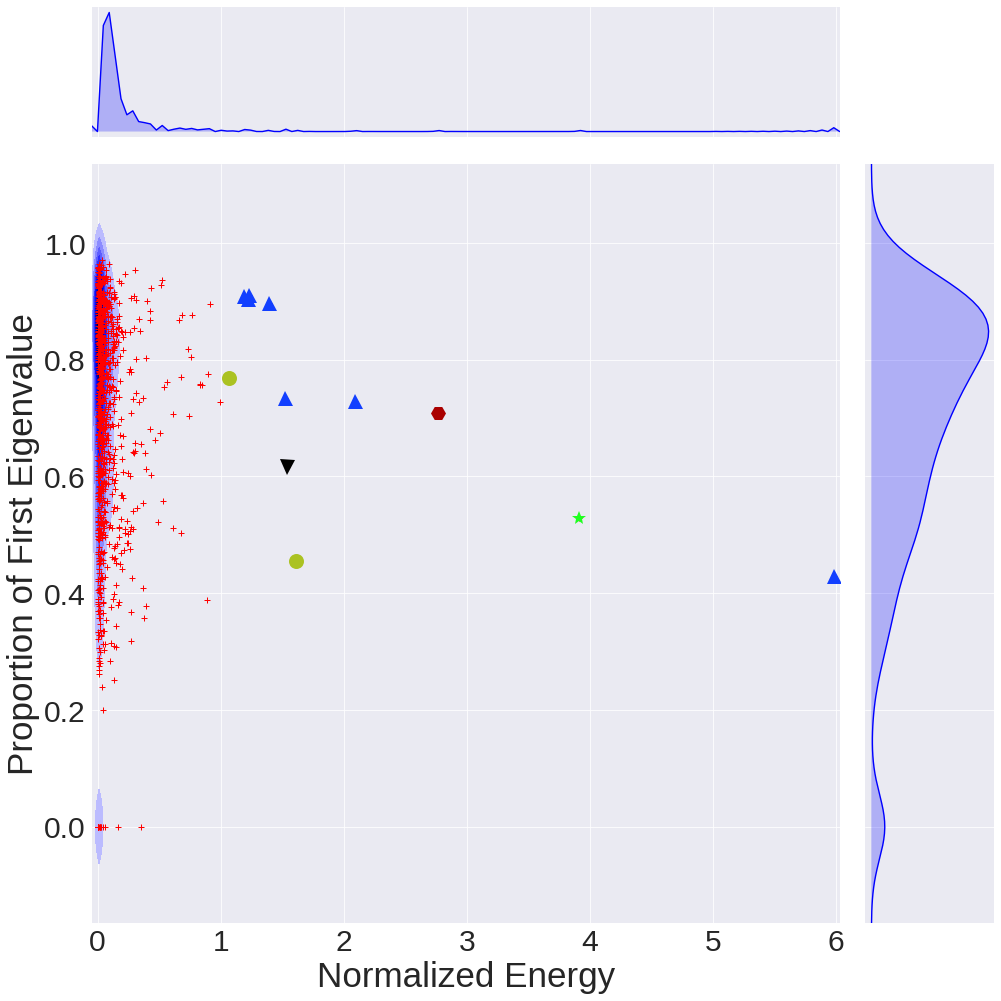}}\label{fig:phase3}}
\qquad    \subfloat[Zoom of the all datasets phase diagram ]{{\includegraphics[width=6.5cm]{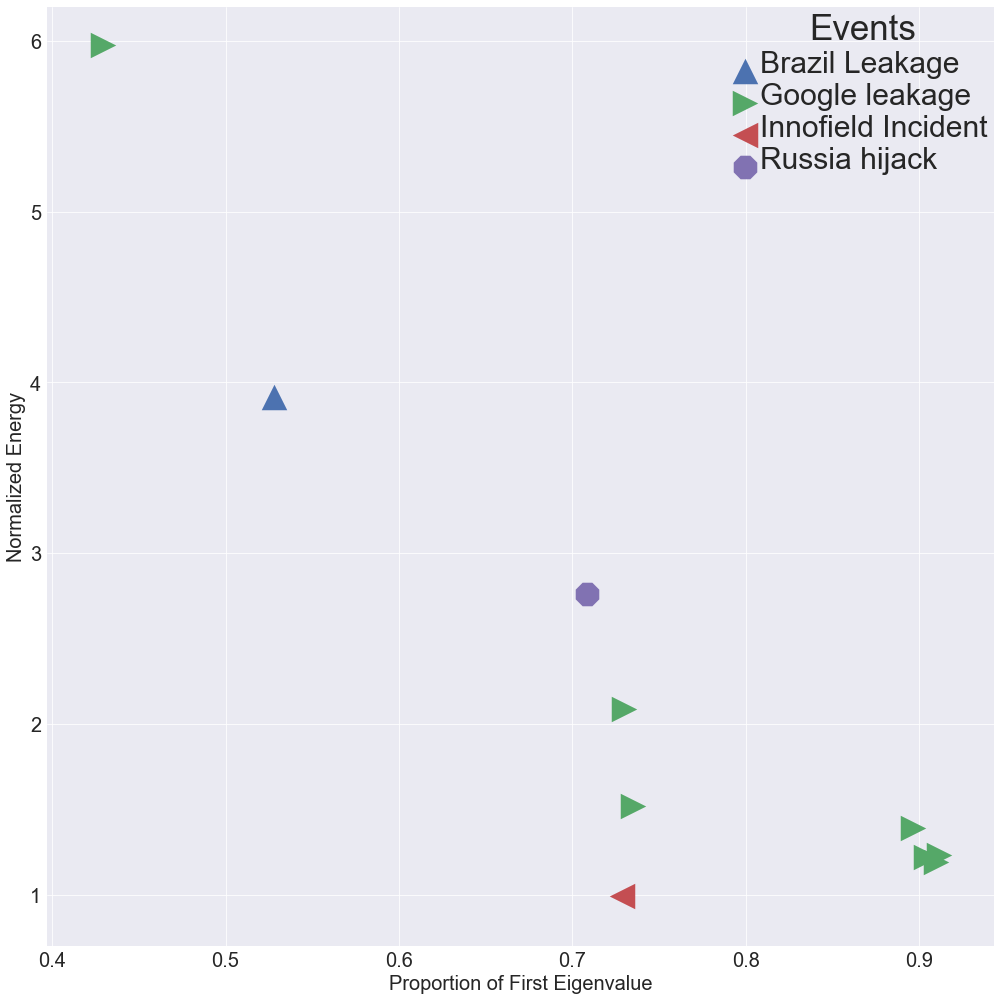} }    \label{fig:phase4}}%
\caption{Phase diagram obtained with all datasets described in Table \ref{tab:majorevents}}
\label{fig:phasediagram}
\end{figure*}
In this section we will describe outcomes of applying the AS graph monitoring approach over traces containing  well-documented large-scale events. We have selected a corpus of diverse incidents whose influences were variable (from a local event  as a local prefix hijacking to global events such as an important black-out). We pick 8 events summarized in table \ref{tab:majorevents}. There have been 4 major BGP events in 2016 and  2017 that get reported in major blogs monitoring BGP events like the BGPMON blog \cite{BGPMON}. All these events are reported in table \ref{tab:majorevents} as "large" events. We add to the table some datasets representing periods without any major reported events. 

We show in Fig. \ref{fig:phasediagram} the phase diagram resulting from combining all datasets in Table \ref{tab:majorevents}. We show the full phase diagram in Fig. \ref{fig:phase3} and a zoom on the values with normalized energy larger than 1 in Fig. \ref{fig:phase4}. As can be seen all major events that have been reported in the major BGP blogs appear in the zoom region. Meaning that all such events generate large spike in their normalized energy.  This observation confirms that the large-scale event detector spots correctly all large-scale events. A more detailed analysis show that all events are detected within 2 one minute AS level graph snapshot, except from the "2008 cable cut" dataset which does not appear in the large-scale event detection region. Nonetheless, we saw for this dataset, several points with large energy (around 0.85) but high (around 0.8) value of $\gamma^{-1}_k$ that were compatible with large local changes in the underlying geometry. A deeper analysis of the $\Delta$ matrix at that times shows a decrease in curvature along with a balancing curvature increase relative to only one or two landmarks. These observations are consistent with reports (see \cite{cable}) that stated that the cable cut was not reported until several day later as its effect was not really felt globally. 

\subsection{Calibrating the large event detector} 
In Fig. \ref{fig:phasediagram} one can see a clear threshold for detecting large-scale event at $\overline{\|\Delta^k\|_F}>1$. However, setting this threshold needs a little more motivation. 
The $\Delta^k$ matrix can be considered as a random matrix with values distributed between 1 and -1. In \cite{singular_value}, Rudelson {\em et al.} derived, through an  estimate of the smallest and largest singular value,  an asymptotic bound for the normalized Frobenius norm of a random sub-gaussian  matrix. Through this estimate for large enough $N$,  an $L \times N$ matrix $A$ with independent and identically distributed sub-gaussian entries with mean 0 and variance 1, the normalized Frobenius norm  will be distributed following a $\chi^2$ distribution with 1 degree of freedom. In our case, the curvature values in $\delta^k$ are bounded, $-2\le \delta^k_{ij}\le 2$ and therefore there are sub-gaussian and the theoretical conditions are valid. 

Now if the entries of the $\delta^k$ matrix have a mean 0 and a variance equal to $\alpha$, one can expect that $\frac{\overline{\|\Delta^k\|_F}}{\alpha}$ will be distributed following a $\chi^2$ distribution with 1 degree of freedom. We tested if the empirical distribution of $\overline{\|\Delta^k\|_F}$ is following a $\chi^2$ distribution with 1 degree of freedom. It is a well known property of $\chi^2$  distribution with $\nu$ degree of freedom that it is equivalent with a Gamma distribution with parameters $(\frac{\nu}{2},2)$. We fitted a Gamma distribution to the all $\overline{\|\Delta^k\|_F}$ values gathered over all datasets and we estimated the parameter to be $(a=0.49, b=2.3)$ that is fully compatible with a $\chi^2$ distribution with 1 degree of freedom showing that the theoretical asymptotic distribution seems to hold in practice.

This asymptotic distribution gives us a way to derive the decision threshold for a large event. By  setting a decision threshold at $T$, {\em i.e.}, if $\overline{\|\Delta^k\|_F}>T$ a large event is detected, the probability of false alarm will be $\mathcal{E}_{\chi^2}(\frac{T}{\alpha})$, where $\mathcal{E}_{\chi^2}(.)$ is the complementary cumulative distribution function of the $\chi^2$ distribution with 1 degree of freedom.  

Empirically over all $\Delta$ matrices we gathered, the elements in the matrix have mean $8.6\times 10^{-5}$ and variance $3.5\time 10^{-2}$, meaning that with a threshold $T=1$ the probability of false alarm, {\em i.e.} wrongly deciding that an observed spike is a large-scale event while it was not, is less than $1.36 \times 10^{-7}$ which gives in average a false alarm each 13,5 years (assuming one snapshot by minute). Interestingly even with such a high threshold, we were able to detect all large-scale events reported during the past 3 years without any false alarm. As we achieved a detection rate of 1 without false alarms, we did not show the ROC curve \cite{ROC} as the result would be trivial.

\subsection{Interpretation of large curvature changes}
\begin{figure}[h]
\includegraphics[width=7cm]{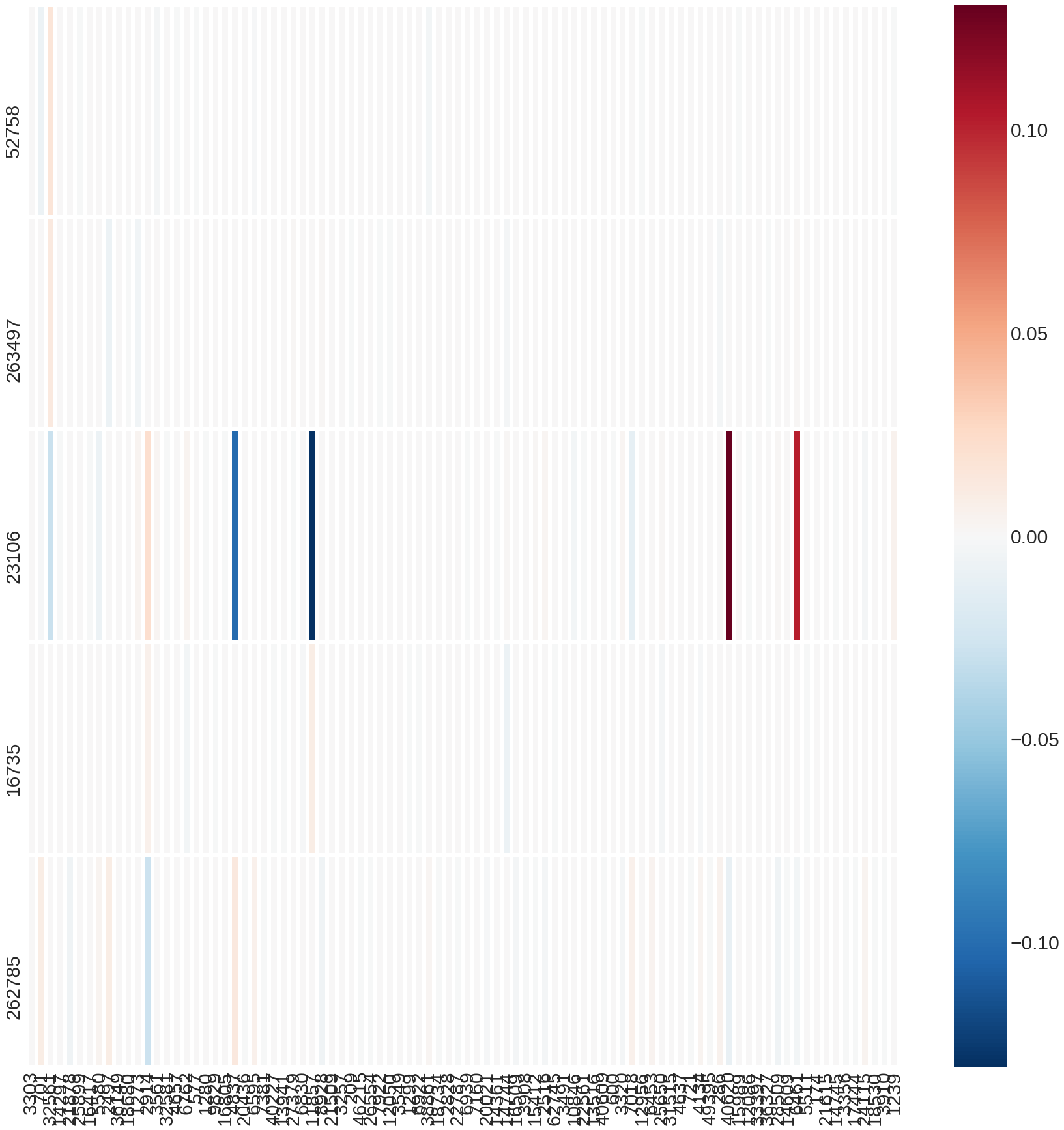}
    \caption{Difference between the transport plan}
    \label{fig:optimal_plan}

\end{figure}

In previous sections we described how to monitor and detect large change in the geometry of AS level graphs. When such an event happens, it is desirable to have some ideas about the source of the problem and to implement a root cause analysis. Interestingly the optimal transport plan $\theta^*$ that is derived during the calculation of Ollivier-Ricci curvature (see Eq. \ref{eq:optTransport}), gives a lot of insights about the causes of the curvature changes. The optimal transport plan $\theta^*(\mu_x, \nu_y)$ can be represented as a matrix where each row is relative to a neighbor of the source $x$ and each column is relative to a neighbor of the destination $y$. One can therefore compare for 2 given nodes $x$ and $y$ this matrix before and after an large-scale event. We show in Fig. \ref{fig:optimal_plan} such a comparison using the difference between two optimal transport plans for the node that have seen the largest curvature change during a large-scale event. The Figure shows clearly that the mass at ASN 23106 that used to be transferred mainly through ASNs 4637 and 8928 (the dark blue elements in Fig. \ref{fig:optimal_plan}) are transferred after the large-scale event by ASNs 6461 and 40620 (the dark red elements in Fig. \ref{fig:optimal_plan})




\section{Related Works}


\noindent \textbf{Anomaly detection of BGP feeds}  Many attempts have been made to detect anomalous Internet events through dissecting BGP updates and tables. A complete and recent survey of BGP Anomaly detection techniques is provided in \cite{BGPsurvey}. Among all these work \cite{Labovitz} used Fourier analysis tools to analyze routing update rates and to relate them to BGP anomalies. Mei {\em et al.} developed BAlet, a wavelet based approach that cluster fast changing  BGP feeds to identify possible source of large-scale anomalies \cite{4575169}. Wavelet Transforms were also used in the BGP-lens \cite{Prakash} to analyze the rate of updates of per prefix. Other statistical approaches like PCA have also been used \cite{Huang} over BGP update rate. 

Several approaches used machine learning techniques to detect link failure or other direct anomalies like BGP Hijacks or indirect one caused by routing overload generated by work propagation for example \cite{Li,Lutu}. However, almost all of them leverage the rate of BGP updates (see for more details \cite{BGPsurvey}).


Gao {\em et al.} \cite{Gao:2001:SIR:504611.504612} infers the non-disclosed policies between ASes by investigating their preferred routes. An "anomaly" is detected when the chosen path does not correspond to the expected policy. This approach is very localized and cannot be extended to the whole graph as it needs one of the monitored ASes to be affected by the "anomaly" for the detection to happen. Moreover, inferring non-disclosed policies might be hard specially with malicious actors. 

While the above approaches target detecting BGP anomalies, they are not dealing with the topic of this paper that is graph monitoring and detecting large-scale event. Moreover, all the above approaches are suffering from the self-similar and heavy tailed structure of update rates caused by BGP session resets that are known to result into large BGP exchanges and inflated update rates. They also do not provide a way to detect further the location and to interpret the anomalies, {\em i.e.}, to infer the cause relative to AS graph change to the observed surge in update rates.

\noindent \textbf{Large-scale events detection}: The previous described approaches were not targeting specifically large-scale events. Comarela {\em et al.} \cite{Comarela} similarly to us targeted large-scale. They aggregate next-hop changes into a metric tensor. By looking at the concentration of elements within the tensor, they proposed a large-scale events detection system. However, they use a daily granularity, which is not satisfactory for small time scale detection. The analysis also heavily relies on hop-count changes that is not a stable enough metric to understand the real evolution of the network as we discussed above (see section \ref{sec:newevent}). In comparison, we provide an almost real-time event detection that is detecting a larger set of events that the one that  hop-count could detect. Chen {\em et al.} \cite{MengChen2015}, extended this paper and developed the LBE, a metric that aims into locating large-scale events. However, LBE depends mainly on the number of BGP updates happening between two detection time and not on the nature of the update.
In \cite{Ku}, PCA is used to separate updates triggered by distinct underlying events. This separation simplifies the root cause analysis but remains highly sensitive to false alarms. 


\noindent \textbf{Traffic volumes anomaly detection} There exists a large literature on anomaly detection in traffic volumes, {\em e.g.}, \cite{soule2005} \cite{lakhina}. Time series analysis approaches are the most widely spread analytic tool used in the context of anomaly detection. Those approaches rely on the definition of a Linear Time independent (LTI) state space synthetic stochastic  model  capturing observed temporal and spatial correlations between observed traffic volumes during "normal" periods. The tracking of this normal model can be done with a Kalmann Filter \cite{soule2005} or without it \cite{lakhina}. 

Our approach in this paper differs mainly from the above method by the fact that we are analyzing the graph, a discrete object, rather than a quantitative metric like traffic volume. One can indeed extend method applicable to volume anomaly detection to the normalized energy time series we have developed in this paper. But the difference between large-scale event and local drift is so large in our observations that a simple thresholding was shown to be sufficient.

\noindent \textbf{Ricci curvature on graphs} Recent works in Riemannian Geometry from Gromov \cite{Gromov} and Perelman \cite{Perelman}, were extended in the past decade to discrete objects and graphs. This extension use intensively the optimal transport approach developed by Lott and Viliani \cite{Lott}. Ollivier developed in\cite{Ollivier:2009} a tractable formula to compute the Ricci curvature over graphs. This attracted a lot of interest in the research community in machine learning and bioinformatics \cite{Whidden}\cite{Pouryahya}\cite{Ache}.
In \cite{Weber}, Weber {\em et al.} presented an approach to obtain a geometric description of highly dynamic networks. They use a different approach of discretization of the Ricci curvature elaborated by Forman \cite{Forman2003}. The relationship between the two approaches have been explained in \cite{Samal}. Forman's curvature depends on a  topological description of the Ricci and is defined with respect to polygonal meshes while Ollivier's curvature is centered around the propagation of Markov Chains and optimal transports characterization of the space. Our assumption is that the AS graph evolves following a latent stochastic process and therefore it is natural for us to favor Ollivier's description.  

\section{Conclusion}
\label{sec:conclusion}
We have presented a new graph embedding technique for monitoring large scale networks.  Our method relies on the curvature, a geometric property linking local scale to global properties. The Ricci curvature that has been initially defined in continuous manifold is extended to graph through the Ollivier-Ricci curvature. We use the Ollivier-Ricci curvature to embed the AS level graph into a $K$-dimensional space. In the embedded space we evaluate the variations of the projection of nodes that have seen a BGP update. This enables us to define a new monitoring system for tracking and detecting large scale events that are changing the overall topology of Internet. We validate the monitoring system over BGP feeds gathered at 20 collectors and we were able to detect in less than 2 minutes all major BGP events that have made the headlines in BGP monitoring blogs. 

The described methodology can be easily extended to any type of complex networks and is attractive for purposes beyond the one mentioned in this paper, such as social network analysis, intra and inter protein networks, {\em etc}. 

One of the main challenge remains to improve the time required for the computation of the optimal transport. This is a very active topic in the machine learning community, see \cite{deep} \cite{solomon}. This would allow the methodology to be deployed live and will provide a complementary monitoring tool for Network Operating Centers. 

\section{appendix}
\subsection{Landmark Selection algorithms}
One obvious way of selecting landmarks is to use the BGP collecting points. But these trivial landmarks will be biased. Nevertheless, collection points have a major advantage as they are accessing an actual BGP feed while for other choice of landmark the AS level topology should be inferred from any other point.

In order to avoid the bias of collection points and to choose suitable landmarks, several approaches are possible. The random choice is appealing because of its simplicity, with the added benefits of security as an attacker may not know which parts of the network he should avoid perturbing to remain undetected. However, random choice in graphs is known to be biased and can result into an over-representation of certain parts of the network. 

Another approach leverages the vertex position in the graph.  Vertices with high centrality are potentially good landmarks as a large number of shortest paths go through them.
Similarly, vertices with the highest degrees are also good candidates for being landmarks. 

Finally, in the context of AS level graphs, tier-1 and tier-2 ASes are another set of candidates. Indeed these vertices are also closely related to the high centrality and high degree vertices. 

All these above approaches besides the random approach have a concentration bias, {\em i.e.}, only central part of the graph where most paths converge is accounted \cite{Lee}. Moreover, picking highly connected vertices implies more expensive computations for calculating the optimal transport distance. Curvature to these central points are also less sensitive to the effects of events.

 In order to deal with these issues, we develop a sampling method to pick good landmarks. To understand intuitively properties of a good set of landmarks, we should go back to the definition of curvature. Let's first look at the sphere that has a positive curvature. By choosing only two landmarks at opposites poles of the sphere and by monitoring the geodesics from these two poles one can detect any changes happening on the sphere surface. While for negative and flat curvature shapes we need an infinity of landmarks to cover all the surface. In general, the geometry resulting from a real graph cannot be simply interpreted as a sphere. One can assume the graph and its underlying geometry is divided into flat valleys or districts separated by high mountains. Good landmarks should still be positioned in such a way that they have large positive curvature in between them in order to have diverging geodesics and not be positioned in the same valleys or district. We leverage this intuition and inspired by the connection made in \cite{Ollivier:2009} between Markov Chain, and curvature, we use a random walk approach to find a satisfying set of landmarks, positioned in with high positive curvature place, far apart from each other to ensure they are not in the same region of the graph and randomly disseminated through the graph to avoid a collusion with an attacker .
 
However, deriving curvature to all other vertices of the network is too complex to be implemented in practice.  A necessary condition for a vertex in the graph to have high curvature is to have a relatively large number of neighbors. One can therefore use the landmark set diversity as defined in Eq. \ref{Eq:S1} as a simple metric to assess possibility of high curvature.

\begin{algorithm}
 \caption{MCMC landmark selection algorithm}
 \label{alg:random_walk}
 \begin{algorithmic}[1]
 \Require
 \Statex Initial Sample $S$
 \Statex Quality Score $P(\cdot)$
 \Statex number of iterations $iter$
 \Ensure
 \Statex Result Sample $R$
 \Statex
 \Procedure{Sampling}{}
 \State $R \gets S$
 \For{ $i$ := 1 to $iter$ }
 \State Select vertex $v \in S$ randomly
 \State  $C \gets (N(v) - S) \cup \{v\}$
 \State Select vertex $w \in C$ randomly
 \State $C' \gets (N(w) - S) \cup \{v\}$
 \State Select $\alpha$ randomly from [0,1]
 \Statex
 \State $S' = (S - \{v\}) \cup \{w\}$
 \State $g(S \to S') = \frac{1}{|S|} \frac{1}{|C|}$
 \State $g(S' \to S) = \frac{1}{|S'|} \frac{1}{|C'|}$
 \State $A(S \to S') = \frac{P(S')}{P(S)} \frac{g(S' \to S)}{g(S \to S')} = \frac{P(S')}{P(S)} \frac{|C|}{|C'|}$
 \Statex
 \If{$\alpha < A(S \to S')$}
 \State $S \gets S'$
 \If{$P(S) > P(R)$}
 \State $R \gets S$
 \EndIf
 \EndIf
 \EndFor
 \EndProcedure
 \end{algorithmic}
 \end{algorithm}

Based on the above considerations, we have developed a  Metropolis-Hasting landmarks selection mechanism, we name \emph{lazy random walk}, that will select landmarks that validate the above constraints. The Metropolis-Hasting algorithm consists
of randomly perturbing a sample set according to a proposal distribution $g(S \to S')$. The distribution $g(S \to S')$  gives the probability of changing a sample S to a sample S'. We  accepts or rejects a new sample set $S'$ based on a quality score $\mathbb{P}(\cdot)$ and a variable $\alpha$. The sampling process consists of first choosing randomly among $L$  candidate landmarks, one vertex $v$ as a candidate for being exchanged with another vertex $w$, {\em i.e.}, $S \to S'=(S-\{v\})\cup \{w\}$. The vertex $w$ is chosen through a random walk among the neighbor of $v$ that are not in the set $S$ and $v$ itself. The sampling probabilities are $g(S \to S') = \frac{1}{|C|}$ and $g(S' \to S) = \frac{1}{|C'|}$, where $C=(N(v)-S)\cup{v}$ is the set of neighbors of $v$ that are not in $S$ plus the vertex $v$ and $C'=(N(w)-S)\cup{v}$ is the set of neighbors of $w$ beside $v$ that are not in $S$. This sampling mechanism ensures that landmark candidate have not too similar neighborhoods and therefore are in different districts of the graph.

Now let's define the score function $\mathbb{P}$. As explained before, the landmarks should be chosen in such a way that they have large positive curvature in between each other. It is however infeasible to derive for each step of the Hasting-Metropolis the $K^2$ curvatures values between  $K$ landmarks. We therefore resort to a simpler heuristic. We define $G_{T_{1,2}}(S)$ as being the set of non-intersecting shortest path from candidate landmarks in $S$ to the AS identified as Tier 1 and Tier 2. By non-intersecting, we mean that there exist no points in common between these twi shortest path. The Tier-1 and Tier-2 vertices are chosen because we already know that these vertices will seat in the center of the network, {\em i.e.}, around the equator for a spherical geometry. Now, by comparing the size of the two sets $G_{T_{1,2}}(S)$ and $G_{T_{1,2}}(S')$, one can assess which of the landmark sets $S$ or $S'$ have the largest positive curvature to the center of the network. This means that we can choose the scoring function as :
$$
P_{\text{updated}}(S) = {\|G_{T_{1,2}}(S)\|}
$$
With the above-defined function, we can use the classical Hasting-Metropolis rejection scheme described in algorithm \ref{alg:random_walk} and implement the sampling.

\bibliographystyle{ACM-Reference-Format}
\bibliography{references}
\end{document}